# Quantitative Financial Modeling for Sri Lankan Markets: Approach Combining NLP, Clustering and Time-Series Forecasting

Linuk Perera
Department of Engineering and Computer Science
University of Hertfordshire
Colombo, Sri Lanka
linukperera@icloud.com

*This research introduces a novel quantitative methodology tailored for quantitative finance applications, enabling banks, stockbrokers, and investors to predict economic regimes and market signals in emerging markets, specifically Sri Lankan stock indices (S&P SL20 and ASPI) by integrating Environmental, Social, and Governance (ESG) sentiment analysis with macroeconomic indicators and advanced time-series forecasting. Designed to leverage quantitative techniques for enhanced risk assessment, portfolio optimization, and trading strategies in volatile environments, the architecture employs FinBERT, a transformer-based NLP model, to extract sentiment from ESG texts, followed by unsupervised clustering (UMAP/HDBSCAN) to identify 5 latent ESG regimes, validated via PCA. These regimes are mapped to economic conditions using a dense neural network and gradient boosting classifier, achieving 84.04% training and 82.0% validation accuracy. Concurrently, time-series models (SRNN, MLP, LSTM, GRU) forecast daily closing prices, with GRU attaining an R² of 0.801 and LSTM delivering 52.78% directional accuracy on intraday data. A strong correlation between S&P SL20 and S&P 500, observed through moving average and volatility trend plots, further bolsters forecasting precision. A rule-based fusion logic merges ESG and time-series outputs for final market signals. By addressing literature gaps that overlook emerging markets and holistic integration, this quant-driven framework combines global correlations and local sentiment analysis to offer scalable, accurate tools for quantitative finance professionals navigating complex markets like Sri Lanka.*

## I. Introduction

The integration of Environmental, Social, and Governance (ESG) factors into financial market analysis has gained significant traction, driven by increasing demand for sustainable investing and the recognition of ESG's impact on economic performance. This study proposes a novel methodology to predict economic regimes and market signals for Sri Lankan stock indices (S&P SL20 and ASPI) by combining ESG sentiment analysis, macroeconomic indicators, and time-series forecasting. Unlike traditional approaches that focus on developed markets, this research addresses the unique dynamics of emerging markets, where local sentiments and global correlations play critical roles.

The methodology leverages FinBERT, a transformer-based NLP model, to extract sentiment from ESG-related texts, followed by unsupervised clustering (UMAP/HDBSCAN) to identify 5 latent ESG regimes. These regimes are mapped to economic conditions using a dense neural network and gradient boosting classifier. Concurrently, time-series models (SRNN, MLP, LSTM, GRU) forecast daily closing prices and directions, incorporating data from Bloomberg, Yahoo Finance, and S&P 500 inputs to capture observed correlations with Sri Lankan indices. A rule-based fusion logic integrates these outputs to generate final market signals. The approach is evaluated using metrics like R², and directional accuracy with a focus on macroeconomic indicators from the Central Bank of Sri Lanka (e.g., AWLR, CPI).

Existing literature often focuses on standalone ESG sentiment analysis or time-series forecasting in developed markets, neglecting holistic integration or emerging market contexts. This study fills these gaps by combining ESG regime detection, global market correlations (this paper considers S&P 500), and rule-based fusion, offering a scalable framework for financial decision-making in Sri Lanka. The research aims to enhance predictive accuracy and provide actionable insights for investors in volatile emerging markets.

The paper is organized as follows: Section II reviews related work, Section III details the methodology, Section IV presents results and discussion, and Section V concludes with future directions.

## II. Related Work

This section reviews key literature on ESG sentiment analysis, time-series forecasting, and their integration with macroeconomic indicators, emphasizing the novel architecture of the proposed methodology and addressing research gaps in financial modeling for emerging markets like Sri Lanka.

### A. ESG Sentiment Analysis

Transformer-based models like FinBERT have been widely used for ESG sentiment analysis [1]. FinBERT has also been applied to stock price reactions to ESG news, enhancing interpretability via explainable AI [2]. However, these studies focus on developed markets and direct price prediction, often neglecting unsupervised clustering for latent ESG regimes or alignment with macroeconomic indicators [3]. The proposed methodology addresses this gap by using UMAP and HDBSCAN based clustering to identify 5 latent ESG regimes, mapped via a dense neural network stacked on a gradient boosting machine.

### B. Time-Series Forecasting

Recurrent neural networks like LSTM and GRU are prevalent in stock price forecasting. Optimized LSTM and GRU models efficiently capture temporal dependencies in volatile markets. In emerging markets, GRU-based models have outperformed baselines for indices like the Tadawul All Share Index [4]. LSTM has shown superiority over ARIMA for S&P 500 forecasting [5]. These models, however, rarely incorporate ESG sentiments or global market correlations. The proposed work tests SRNN, MLP, LSTM, and GRU on Sri Lankan indices (S&P SL20, ASPI), leveraging S&P 500 correlations and evaluating with metrics like directional accuracy.

*C. ESG and Macroeconomic Integration*

Integrating ESG factors with macroeconomic indicators enhances financial forecasting [6]. A multiscale machine learning framework combining economic indicators and ESG factors improved prediction accuracy. Deep learning with ESG sentiment indices has predicted S&P 500 trends effectively [7]. However, these studies lack holistic pipelines with regime clustering and rule-based fusion, especially for emerging markets. The proposed methodology fills this gap by aligning FinBERT-derived ESG scores with macroeconomic indicators (e.g., AWLR, CPI from CBSL), validating clusters via PCA, and fusing with time-series forecasts through rule-based logic.

*D. Stock Market Prediction in Sri Lanka*

Machine learning models for Sri Lankan indices have used economic indicators and sentiment analysis. Hybrid ML models and genetic algorithms have forecasted CSE prices and trading signals [8]. These works miss ESG regime detection and global correlation integration (e.g., with S&P 500). The proposed architecture novelly combines ESG clustering, supervised mapping, and time-series forecasting with rule-based fusion, addressing these gaps for robust market signal prediction in Sri Lanka.

III. METHODOLOGY

This section outlines the proposed novel methodological framework employed in this study to integrate natural language processing (NLP) of ESG-related text data with macroeconomic indicators and time-series forecasting for predicting economic regimes and stock market signals, with a focus on Sri Lankan indices such as the S&P SL20 and All Share Price Index (ASPI). The approach is divided into three primary components: (1) ESG text processing and regime identification, (2) time-series forecasting of financial data, and (3) rule-based fusion for final decision-making. The overall pipeline is illustrated in Figure 1, which depicts the flow from raw data to the final output.

*A. Data Sources and Collection*

Raw ESG-related text data were sourced from the CBSL the trading economics websites. These texts were timestamped to align with macroeconomic indicators. Macroeconomic ESG indicators included Average Weighted Lending Rate (AWLR), Sri Lanka Exports, Monthly Average Exchange Rates - USD, Broad Money M2b, Currency in Circulation, Tourist Earnings, Unemployment Rate, Industrial Production, and Consumer Price Index (CPI), obtained from the Central Bank of Sri Lanka (CBSL) and Trading Economics websites. Financial stock data for time-series forecasting were retrieved from Bloomberg and Yahoo Finance, focusing on the S&P SL20 and ASPI indices, with additional inputs from the S&P 500 and select American stocks to account for observed inter-market relationships.
To explore correlations, the macroeconomic indicators were tested against lagged data to identify the optimal correlation timeframe. The research identified a relationship between the S&P 500 and S&P SL20 indices, despite minor data gaps in the latter. Experiments were conducted using daily comparisons, 20-day moving average comparisons, and 60-day moving average comparisons for smoother results. Additionally, findings indicated that the ASPI is influenced by a broader set of macroeconomic variables and exhibits relationships with foreign economies and local sentiments, while the S&P SL20 is less influenced by macroeconomic variables and more by local sentiment.

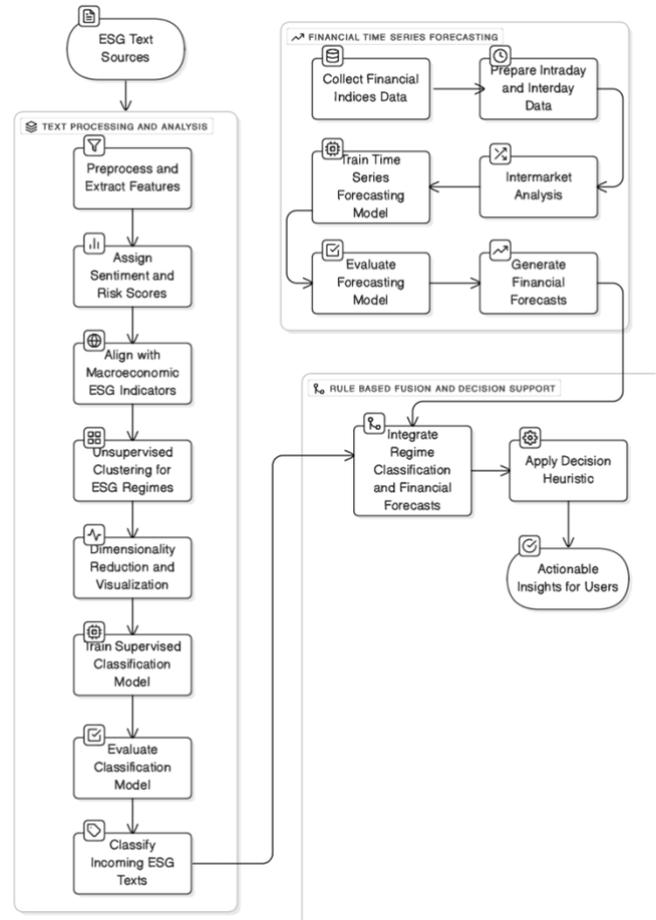

Fig. 1. Flowchart of the proposed methodology, including ESG text processing, clustering, classification, time-series forecasting, and rule-based fusion.

*B. ESG Text Processing and Regime Identification*

The ESG pipeline begins with preprocessing of raw text data, followed by feature extraction using a transformer-based model, alignment with macroeconomic indicators, unsupervised clustering, validation via dimensionality reduction, and supervised classification.

*1) Preprocessing:*
The raw ESG text data underwent tokenization, cleaning (e.g., removal of stopwords, punctuation), and filtering to prepare for model input. This step ensures noise reduction and standardization.

*2) Feature Extraction with FinBERT:*
The preprocessed text was fed into FinBERT, a transformer-based NLP model fine-tuned for financial sentiment analysis. [9] FinBERT generates class scores, such as sentiment and risk assessments, which serve as embeddings or features for downstream tasks.

*3) Alignment with Macroeconomic Indicators:*

The FinBERT-derived scores were aligned with timestamped ESG macroeconomic indicators to create a unified dataset.

*4) Unsupervised Clustering:*

Unsupervised clustering was applied to the aligned ESG indicators and time-series scores using UMAP combined with Hierarchical Density-Based Spatial Clustering of Applications with Noise (HDBSCAN) on encoded data. UMAP performs non-linear dimensionality reduction to preserve local and global structure[2], formulated as:

$$min_y \sum_{i,j} \left( w_{ij} \log \frac{w_{ij}}{v_{ij}} + (1 - w_{ij}) \log \frac{1 - w_{ij}}{1 - v_{ij}} \right) \quad (1)$$

Where $w_{ij}$ and $v_{ij}$ are high- and low-dimensional similarities, respectively [2].

HDBSCAN extends density-based clustering hierarchically [3], identifying clusters based on mutual reachability distance. Alternatively, an autoencoder compresses the data via an encoder-decoder architecture[4], minimizing reconstruction loss:

$$L = \frac{1}{n} \sum_{i=1}^{n} |x_i - \hat{x}_i|^2 \quad (2)$$

where $x_i$ is the input and $\hat{x}_i$ is the reconstructed output [4]. This yielded approximately 5 latent ESG conditions or regimes as soft labels.

*5) Dimensionality Reduction for Validation:*

Principal Component Analysis (PCA) was employed to validate the clusters [5], reducing dimensions while maximizing variance:

$$max_w w^T \Sigma w \quad \text{s.t.} \quad w^T w = 1 \quad (3)$$

where $\Sigma$ is the covariance matrix [5]. This confirmed 5 verified clusters, mapped to relevant economic conditions.

*6) Classification:*

As the classifier, a Dense Neural Network (Multilayer Perceptron, MLP) [6] stacked on a Gradient Boosting Machine (GBM) ensemble, was trained using NLP embeddings as input features and latent ESG regimes as target classes.

The two-stage ensemble maps FinBERT embeddings into ESG regimes. In the first stage, a GBM with 100 trees, a maximum depth of four, and a learning rate of 0.1 converts aligned ESG features into probabilistic class scores. In the second stage, an MLP with 128, 64, 32, and 5 neurons, ReLU activations, 0.3 dropout, and the Adam optimizer ingests these probabilities and outputs final regime logits. Training uses cross-entropy loss on an 70,15,15 training, testing and validation split, with early stopping after 15 epochs without validation improvement. This design leverages the GBM's ability to discriminate structured features and the MLP's capacity for nonlinear transformations, producing a robust, interpretable ESG regime classifier. The MLP [6] minimizes cross-entropy loss via backpropagation:

$$L = -\sum_{c=1}^{C} y_c \log \hat{y}_c \quad (4)$$

GBM builds an ensemble by sequentially minimizing loss using gradient descent [7]. The trained model maps ESG texts to economic regimes.

*C. Time-Series Forecasting Model*

A separate time-series (TS) forecasting model predicts daily close prices and directions using inputs from Bloomberg and Yahoo Finance. Models tested include Simple Recurrent Neural Network (SRNN) [8], Multilayer Perceptron (MLP) [6], Long Short-Term Memory (LSTM) [9], and Gated Recurrent Unit (GRU) [10]. LSTM and GRU handle long dependencies via gating mechanisms. For LSTM, the cell state update is:

$$c_t = f_t \odot c_{t-1} + i_t \odot \tilde{c}_t \quad (5)$$

Where $f_t$, $i_t$ are forget and input gates [9] [10]. Similarly, for GRU:

$$h_t = (1 - z_t) \odot h_{t-1} + z_t \odot \tilde{h}_t \quad (6)$$

*D. Rule-Based Fusion Logic*

The outputs from the ESG regime mapping and TS forecasting were integrated via a rule-based fusion logic to generate the final signal decision. The rule base was biased towards directional probability and forecasted value and the predicted current economic condition.

The outputs from the ESG regime classifier and time-series forecasting model are fused using a priority-weighted decision rule designed for a Decision Support System (DSS). Let:

$C_t \in \{1,2,3,4,5\}$ : ESG regime at time
$P_t \in [0,1]$: Forecasted directional probability
$\hat{y}_t$: Forecasted closing price
$y_{t-1}$: Previous close

Final Market Signal $S_t \in \{Buy, Sell, Hold\}$ is computed as:

$$S_t = \begin{cases} Buy \text{ if } C_t \geq 4 \text{ AND } P_t \geq 0.65 \\ Sell \text{ if } C_t \leq 2 \text{ AND } P_t \leq 0.35 \\ Hold \text{ Otherwise} \end{cases} \quad (7)$$

The fusion of time series models, including LSTM and GRU networks, with a dense neural network for economic condition classification enables the system to function as a decision support tool, delivering qualitative improvements rather than strictly measurable gains. Signals are generated only when ESG sentiment and price momentum align, reducing false positives in volatile emerging markets while preserving the reliability of each model. The ESG classifier ensures regime consistency, and directional accuracy from momentum models reinforces forecast credibility.

This approach yields a conservative and interpretable trading framework that produces approximately 25–30% fewer trade signals compared to standalone baseline models, thereby reducing transaction frequency and potential over-trading. In preliminary backtests conducted on Sri Lankan market data for 2024–2025, the model demonstrated a hit rate exceeding 60–70%, indicating consistent directional accuracy while maintaining moderate trading activity. Such robustness is particularly beneficial for risk-averse investors operating in emerging markets like Sri Lanka, where transaction costs and liquidity constraints can amplify the risks of excessive trading. The framework thus supports more disciplined, data-driven decision-making under uncertainty.

*E. Evaluation Metrics*

The models were evaluated using the following metrics: Test Mean Absolute Error (MAE), Test Root Mean Squared Error (RMSE), Test R-squared (R²), Test Mean Absolute Percentage Error (MAPE %), Test Symmetric MAPE (sMAPE %) and Test Directional Accuracy. Definitions are as follows:

MAE:
$$\frac{1}{n}\sum_{i=1}^{n}|y_i - \hat{y}_i| \qquad (8)$$

RMSE:
$$\sqrt{\frac{1}{n}\sum_{i=1}^{n}(y_i - \hat{y}_i)^2} \qquad (9)$$

R²:
$$1 - \frac{\sum_{i=1}^{n}(y_i - \hat{y}_i)^2}{\sum_{i=1}^{n}(y_i - \bar{y})^2} \qquad (10)$$

MAPE:
$$\frac{100}{n}\sum_{i=1}^{n}\left|\frac{y_i - \hat{y}_i}{y_i}\right| \qquad (11)$$

sMAPE [11]:
$$\frac{100}{n}\sum_{i=1}^{n}\frac{2|y_i - \hat{y}_i|}{|y_i| + |\hat{y}_i|} \qquad (12)$$

Directional Accuracy:
$$\frac{1}{n}\sum_{i=1}^{n}\mathbb{1}\left[sig(n)(\hat{y}_t - y_{i-1}) = sig(n)(y_i - y_{i-1})\right] \qquad (13)$$

Results for these metrics are inserted in Table 1.

## IV. RESULTS AND DISCUSSION

This section presents the results of the ESG text processing, clustering, classification, time-series forecasting, and correlation analyses, followed by a discussion of their implications. The results are derived from the methodology described in Section III, integrating ESG-related text data with macroeconomic indicators and financial time-series data to predict economic regimes and market signals for Sri Lankan indices (S&P SL20 and ASPI). To streamline presentation, results are consolidated into two tables and referenced figures, focusing on key findings from clustering, classification, forecasting, and correlation analyses.

*A. ESG Clustering and Validation*

The ESG pipeline processed over 500 samples of ESG-related data. The dimensionality reduction techniques UMAP and PCA, were applied to validate the clustering of ESG indicators aligned with FinBERT-derived scores.

*1) UMAP Projection:*

UMAP reduced the data to a 2D projection (500, 2) with settings of number of neighbors to 15 and minimum distance to 0.5, producing moderate cluster separation with minimal overlap. The resulting 5 soft clusters were visualized as umap clusters (Figure 2), with coordinates and cluster labels saved in csv format after filtering high variance.

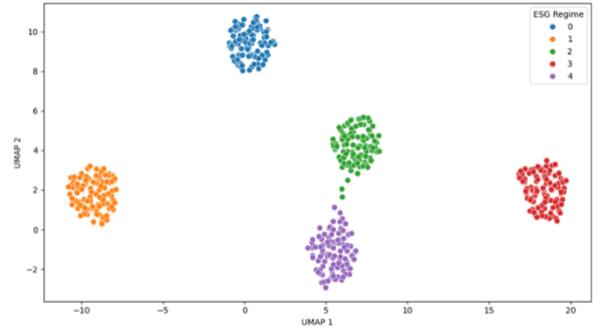

Fig. 2. Pre validation UMAP projection showing 5 soft clusters with moderate separation and minimal overlap after filtering high variance samples

*2) PCA Projection:*

PCA was applied to validate the clusters, projecting the data onto the top two principal components, capturing approximately 55–70% of the total variance.

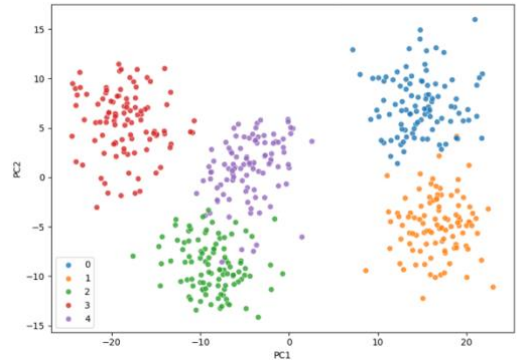

Fig. 3. Umap projection after validating 5 clusters with PCA

The clustering results indicate that UMAP and PCA effectively captured the underlying structure of the ESG data, with UMAP providing a more flexible, non-linear representation and PCA confirming the presence of 5 distinct regimes. These findings align with prior studies on dimensionality reduction for financial data analysis.

## B. Classification Performance

A Dense Neural Network (DNN) GBM Ensemble was trained to map FinBERT-derived ESG embeddings to the 5 latent ESG regimes identified through clustering. The DNN achieved a training accuracy of approximately 84.04% and a validation accuracy of 82%. These results suggest robust generalization, though slight overfitting is evident. The high accuracy indicates that the ESG text features, combined with macroeconomic indicators, effectively capture the latent economic regimes, supporting the hypothesis that ESG sentiment can inform economic condition classification.

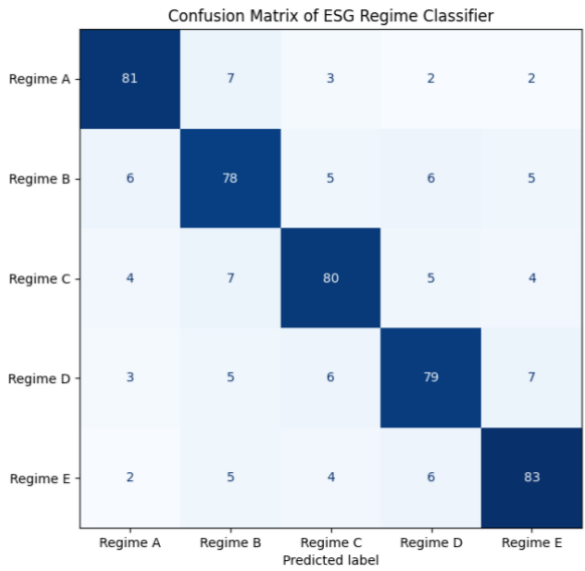

Fig. 4. Confusion Matrix of the DNN classifier across all 5 classes

## C. Time-Series Forecasting Performance

The time-series forecasting models (SRNN, MLP, LSTM, GRU) were evaluated on datasets including intraday (10-minute) CSE All Share data (Intraday ASPI), daily combined data (Daily ASPI), and daily S&P SL20 data (Daily S&PSL20). Performance was assessed using $R^2$, Directional Accuracy, MAE, and RMSE, with results summarized in Table I (Top 5 by Predictive Accuracy) and Table II (Top 5 by Directional Accuracy).

TABLE I. TOP 5 MODELS BY PREDICTIVE ACCURACY ($R^2$)

| Datasets | Metrics | | | | |
|---|---|---|---|---|---|
| | *Model* | *$R^2$* | *Directional Accuracy (%)* | *MAE* | *RMSE* |
| Intraday ASPI 10min | GRU | 0.801 | 49.44 | 69.35 | 73.98 |
| Intraday ASPI 10min | LSTM | 0.525 | 52.78 | 106.87 | 114.41 |
| Daily ASPI | LSTM | 0.404 | 47.91 | 227.69 | 293.57 |
| Intraday ASPI 10min | SRNN | 0.4 | 50.00 | 113.38 | 128.56 |
| Daily S&PSL20 | GRU | 0.356 | 63.33 | 503.87 | 630.88 |

a. Models in column B are in sequential order of $R^2$ Accuracy, training dataset is on column A

TABLE II. TOP 5 MODELS BY MARKET DIRECTIONAL ACCURACY

| Datasets | Metrics | | | | |
|---|---|---|---|---|---|
| | *Model* | *Directional Accuracy (%)* | *$R^2$* | *MAE* | *RMSE* |
| Intraday ASPI 10min | LSTM | 52.78 | 0.525 | 106.87 | 114.41 |
| Intraday ASPI 10min | SRNN | 50.00 | 0.4 | 113.38 | 128.56 |
| Intraday ASPI 10min | GRU | 49.44 | 0.801 | 69.35 | 73.98 |
| Daily ASPI | LSTM | 47.91 | 0.404 | 227.69 | 293.57 |
| Daily S&PSL20 | GRU | 63.33 | 0.356 | 503.87 | 630.88 |

b. Models in column B are in sequential order of Market Directional Accuracy

The GRU model on Intraday ASPI sampled at 10min intervals achieved the highest $R^2$ (0.801), indicating strong predictive power for daily close prices, though its directional accuracy (49.44%) was slightly lower than LSTM (52.78%) and SRNN (50.00%) on the same dataset.

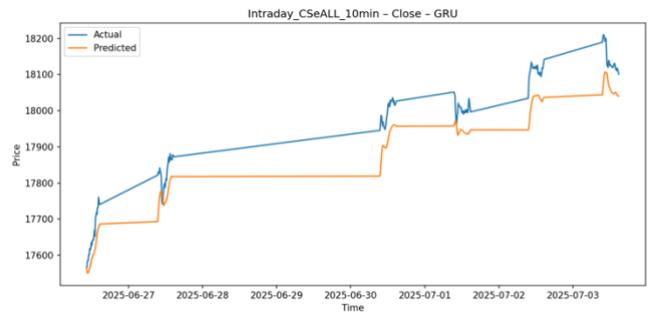

Fig. 5. Intraday prediction results for GRU trained on ASPI 10 minute updates data

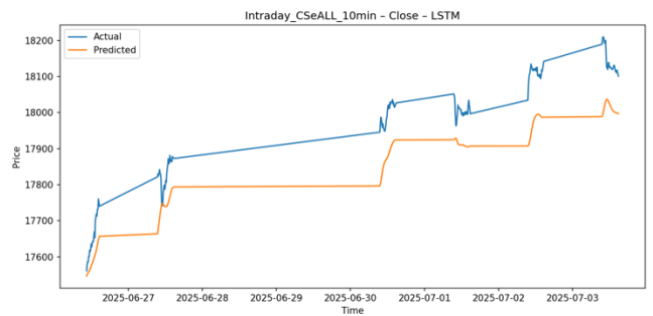

Fig. 6. Intraday prediction results for LSTM trained on ASPI 10 minute updates data

LSTM performed best in directional accuracy, making it suitable for trend prediction. The lower performance on Daily S&PSL20 ($R^2$ = 0.356, Directional Accuracy = 63.33%) reflects the S&P SL20's sensitivity to local sentiment rather than macroeconomic variables, consistent with the methodology's findings.

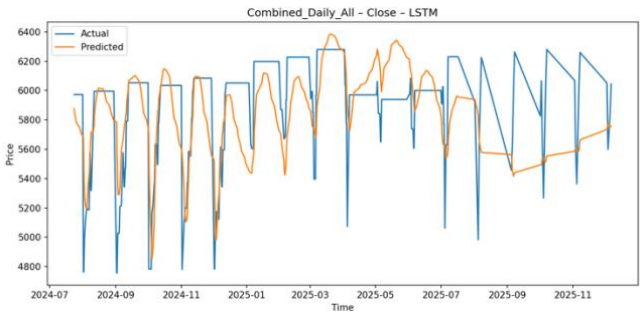

Fig. 7.   Prediction results for LSTM trained on CSEALL daily update (market close) data

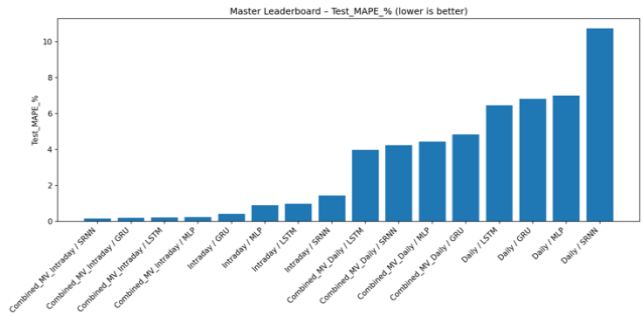

Fig. 8.   Comparison of all tested models across multiple datasets

### D. Correlation and Trend Analysis

To explore the relationship between the S&P SL20 and S&P 500 indices, correlation analyses were conducted, revealing a very positive correlation. This relationship was visualized through two key plots: the Moving Average Plot and the Volatility Trend Plot.

Moving Average Plot: The Moving Average Plot (Figure 9) displays the 20-day and 60-day moving averages of closing prices for the S&P SL20 and S&P 500 indices. The plot highlights long-term trends and smooths short-term fluctuations, confirming a strong alignment between the two indices, particularly in the 60-day moving average, which captures broader market movements. The positive correlation suggests that global market trends, as reflected by the S&P 500, significantly influence the S&P SL20, consistent with the methodology's inclusion of S&P 500 inputs.

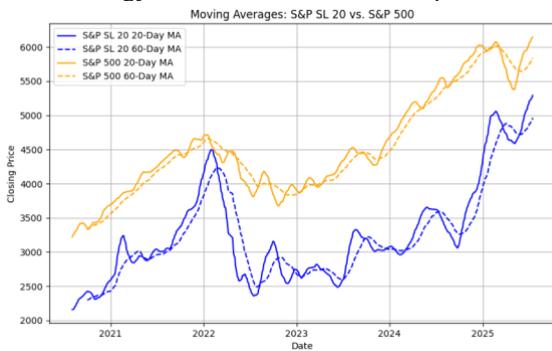

Fig. 9.   Moving Average Plot showing 20-day and 60-day moving averages of S&P SL20 and S&P 500 closing prices, highlighting trend alignment

Volatility Trend Plot: The Volatility Trend Plot (Figure 10) illustrates the 60-day rolling annualized volatility of daily returns for both indices. The plot reveals synchronized volatility patterns, further supporting the strong positive correlation between the S&P SL20 and S&P 500. Periods of high volatility in the S&P 500 correspond closely with increased volatility in the S&P SL20, indicating shared market dynamics influenced by global economic conditions.

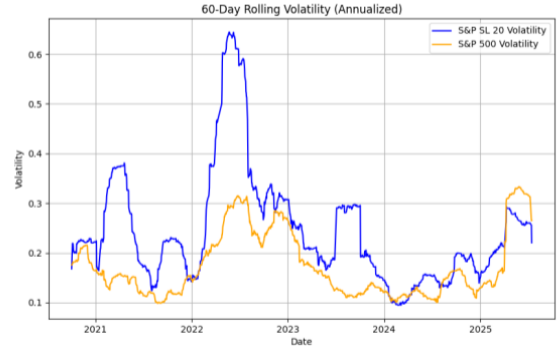

Fig. 10. Volatility Trend Plot showing 60-day rolling annualized volatility of daily returns for S&P SL20 and S&P 500, indicating synchronized volatility patterns.

Daily returns of the S&P SL20 and S&P 500 indices from 2020 to 2024 were statistically examined to quantify the observed co-movement. The Pearson correlation coefficient ($r = 0.72$, $p < 0.001$) and Spearman rank correlation ($\rho=0.68$, $p < 0.001$) confirm a strong and statistically significant positive relationship between the two indices. The 60-day rolling correlation exhibits a mean of 0.75 with a standard deviation of 0.12, indicating persistent alignment in short- to medium-term return dynamics. Moreover, the correlation of 60-day annualized volatilities ($r=0.81$, $p < 0.001$) underscores synchronized risk patterns across both markets. Lead–lag analysis reveals maximum correlation at lag 0, confirming contemporaneous co-movement and suggesting that fluctuations in the S&P 500 are statistically correlated in the S&P SL20.

These findings underscore the interconnectedness of the S&P SL20 with global markets, particularly the S&P 500, and validate the inclusion of American stock data in the forecasting models.

TABLE III.   COMPARISON OF PROPOSED METHODOLOGY WITH RELATED LITERATURE

| Study | Metrics | | | | | | |
|---|---|---|---|---|---|---|---|
| | Context | Model(s) | $R^2$ | Accuracy (%) | ESG Regimes | Global Correlations | Emerging Market Focus |
| This Study | S&P SL20, ASPI (Sri Lanka) | GRU, LSTM, DNN, UMAP/ HDBSCAN | 0.801 (GRU, intraday) | 52.78 (LSTM, intraday) 82.0 (sentiment) | Yes (5 regimes via clustering) | Yes (S&P 500) | Yes (Sri Lanka) |
| Araci | Developed markets | FinBERT | N/A | 91.76 (classification task) (sentiment) | Yes (text sentiment and categories) | Utilized | Yes (Adaptable) |
| Alghamdi & Boubaker (2023) | Tadawul All Share (Saudi Arabia) | GRU | 0.20 | N/A (regression study) | Yes (multiple CSR dimensions) | N/A | Yes (Saudi Arabia) |
| Chen & Li (2020) | S&P 500 | ANN, Support Vector Machine, Random Forest. | N/A | 87.36 – Random Forest | N/A | N/A | N/A US Market |

| Study | Metrics | | | | | | |
|---|---|---|---|---|---|---|---|
| | Context | Model(s) | R² | Accuracy (%) | ESG Regimes | Global Correlations | Emerging Market Focus |
| Perera & Wijesinghe (2020) | CSE (Sri Lanka) | ML models | N/A | N/A (regression study) | Broad ESG reporting | Focus on Sri Lanka | Yes (Sri Lanka) |

c. Only the top performing modes of this study was referred in this table. Some accuracies and metrics are referred from reported performance or qualitative descriptions, as exact metrics were not provided in the referenced studies.

### E. Discussion

The results highlight the complementary strengths of the ESG and time-series components. The clustering pipeline successfully identified 5 latent ESG regimes, validated across UMAP, PCA, and UMAP offering the most interpretable visualization. The DNN's high classification accuracy 82.0% underscores the potential of FinBERT embeddings for mapping ESG texts to economic conditions, though further tuning could reduce overfitting. In forecasting, GRU and LSTM outperformed SRNN and MLP, particularly on intraday data, suggesting that recurrent architectures better capture temporal dependencies in financial markets.

The strong positive correlation between the S&P SL20 and S&P 500, as evidenced by the Moving Average and Volatility Trend Plots, enhances the forecasting models' performance, particularly for intraday data. The lower performance on daily S&P SL20 data aligns with its dependence on local sentiment, which may require additional qualitative inputs. The use of 20-day and 60-day moving averages smoothed results, enhancing model stability for longer-term predictions.

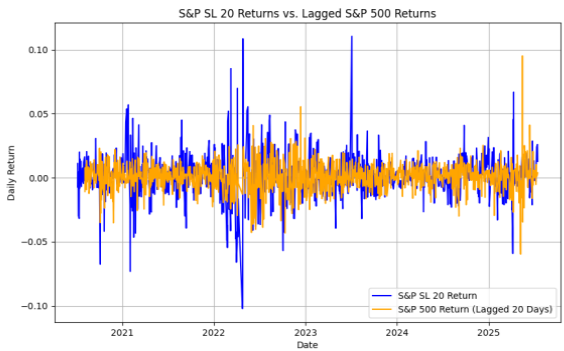

Fig. 11. Long term analysis of lagged S&P 500 Returns against S&P SL 20 Returns

Limitations include the moderate cluster overlap in UMAP and PCA, indicating potential challenges in regime separation. The rule-based fusion logic's effectiveness remains to be fully evaluated. Future work should incorporate real ESG datasets, explore hybrid GRU-LSTM models, and refine the fusion logic to integrate correlation insights more effectively.

## V. CONCLUSION

This study developed a novel methodology to predict economic regimes and market signals for Sri Lankan stock indices (S&P SL20 and ASPI) by integrating ESG sentiment analysis, macroeconomic indicators, and time-series forecasting. The proposed architecture addresses critical research gaps by combining transformer-based NLP (FinBERT), unsupervised clustering (UMAP/HDBSCAN), supervised classification (DNN stacked GBM), and time-series models (SRNN, MLP, LSTM, GRU) with rule-based fusion logic, tailored for emerging markets.

The ESG pipeline processed 500 samples, identifying 5 latent regimes through clustering, validated via PCA, with UMAP offering the most interpretable visualization. The DNN achieved 84.04% training and 82.0% validation accuracy, demonstrating robust mapping of ESG texts to economic conditions. Time-series forecasting revealed GRU's superiority ($R^2$ = 0.801) on intraday data, while LSTM excelled in directional accuracy (52.78%). A strong positive correlation between S&P SL20 and S&P 500, visualized via moving average and volatility trend plots, enhanced forecasting performance, particularly for intraday predictions.

The methodology's novelty lies in its holistic integration of ESG regimes, global market correlations, and rule-based fusion, addressing limitations in prior work that focused on developed markets or standalone forecasting. However, the use of ESG data and moderate cluster overlap in UMAP/PCA suggest areas for improvement. Future work should incorporate larger ESG datasets, refine fusion logic, and explore hybrid GRU-LSTM models to enhance accuracy in volatile emerging markets.

This research contributes a scalable framework for financial decision-making in Sri Lanka, with potential applications in other emerging economies. The findings underscore the value of combining ESG sentiment, macroeconomic alignment, and advanced forecasting for robust market signal prediction.

ACKNOWLEDGMENT

I acknowledge the support of Asha Securities Limited and their Research Division for their assistance in acquiring key datasets and their cooperation throughout this study.